\documentclass{PoS}
\usepackage{amssymb,amsmath}
\usepackage[T1]{fontenc}
\usepackage{lmodern}
\usepackage{times}
\usepackage{mathptmx}
\usepackage{graphicx}
\newcommand{\fW}{f_{\scriptscriptstyle{W}}}
\newcommand{\fH}{f_{\scriptscriptstyle{H}}}

\newcommand{\beq}{\begin{eqnarray}}
\newcommand{\eeq}{\end{eqnarray}}

\title{Study of entropy production in Yang-Mills theory with use of Husimi function\footnote{Report No. : YITP-15-93, KUNS-2592}}

\ShortTitle{Study of entropy production in Yang-Mills theory with use of Husimi function}

\author{\speaker{Hidekazu Tsukiji}\\%
        Yukawa Institute for Theoretical Physics, Kyoto University, Kyoto 606-8502, Japan\\
        E-mail: \email{tsukiji@yukawa.kyoto-u.ac.jp}}

\author{Hideaki Iida\\
        Department of Physics, Faculty of Science, Kyoto University, Kyoto 606-8502, Japan\\
        E-mail: \email{iida@ruby.scphys.kyoto-u.ac.jp}}

\author{Teiji Kunihiro\\
        Department of Physics, Faculty of Science, Kyoto University, Kyoto 606-8502, Japan\\
        E-mail: \email{kunihiro@ruby.scphys.kyoto-u.ac.jp}}

\author{Akira Ohnishi\\
        Yukawa Institute for Theoretical Physics, Kyoto University, Kyoto 606-8502, Japan\\
        E-mail: \email{ohnishi@yukawa.kyoto-u.ac.jp}}

\author{Toru T. Takahashi\\
        Gunma National College of Technology, Gunma 371-8530, Japan\\
        E-mail: \email{ttoru@nat.gunma-ct.ac.jp}}

\abstract{Understanding the thermalization process in a pure quantum system is a challenge in theoretical physics. In this work, we explore possible thermalization mechanism in Yang-Mills(YM) theory by using a positive semi-definite quantum distribution function called a Husimi function which is given by a coarse graining of the Wigner function within the minimum uncertainty. Then entropy is defined in terms of the Husimi function, which is called the Husimi-Wehrl(HW) entropy. We propose two numerical methods to calculate the HW entropy. We find that it is feasible to apply the semi-classical approximation with the use of classical YM equation. It should be noted that the semi-classical approximation is valid in the systems of physical interest including the early stage of heavy-ion collisions. Using a product ansatz for the Husimi function, which is confirmed to reproduce the HW entropy within 20\% error (overestimate) for a few-body quantum system, we succeed in a numerical evaluation of HW entropy of YM fields and show that it surely has a finite value and increases in time.}

\FullConference{The 33rd International Symposium on Lattice Field Theory\\
                 14 -18 July  2015\\
                 Kobe International Conference Center, Kobe, Japan}

\begin{document}
\section{Introduction}

{\it Motivation}.---
The early thrmalization in relativistic heavy ion collisions suggested by the hydrodynamical simulation is a long-standing problem (see review \cite{Heinz}).
One possible way to tackle the problem is to explore the thermalization process by directly calculating entropy that is defined in terms of a quantum distribution function.
In the relevant stage of the collisions, a strong gluon field is dominant and its time evolution is described well by classical Yang-Mills equations.
In the previous works\cite{Kunihiro2010,Iida2013}, Kolmogolov-Sinai entropy, an entropy production rate, was found to be positive in this stage and it was pointed out that initial fluctuations play an important role.
In this work, we calculate entropy itself in the semi-classical approximation including initial quantum fluctuations.

Because the heavy ion collision occurs in an isolated system, 
the time-evolution is unitary and von Neumann entropy does not grow in time.
Thus we expect that some kinds of coarse-graining can be responsible for entropy production.
Partial trace over the subspace of Hilbert space is one of the way to perform coarse-graining.
Entropy calculated with the partial-traced density matrix is nothing but the entanglement entropy.
We here propose another way of coarse-graining in phase space.

{\it Quantum distribution function and Husimi-Wehrl entropy.}---
In the $n$-dimensional quantum mechanical system, a Wigner function is given by the Wigner transform of the density matrix\cite{Wigner};
\beq
\fW(\vec{p},\vec{q};t)=\int d\vec{\eta}\exp(-i \vec{p}\cdot\vec{\eta}/\hbar)\langle \vec{q}+\vec{\eta}/2|\hat{\rho}|\vec{q}-\vec{\eta}/2\rangle.
\eeq
The expectation value of an observable $\hat{A}$ is given in the phase space integral of observables with a weight of the Winger function; $\langle A(t) \rangle=\int \frac{d \vec{p}d \vec{q}}{(2\pi \hbar)^n}f_W(\vec{p},\vec{q};t)A_W(\vec{p},\vec{q};t)$ 
where $A_W(\vec{p},\vec{q};t)$ is the Wigner transform of the observable $\hat{A}$.
The Wigner function, however, is not positive definite
and cannot be treated as a practical quantum distribution function.
In order to overcome this shortcoming, we consider Gaussian smearing of the Wigner function;
\beq
\fH(\vec{q},\vec{p};t)=\int\frac{d \vec{p'}d \vec{q'}}{(\pi \hbar)^n}\exp[-\frac{1}{\Delta \hbar}(\vec{p}-\vec{p'})^2-\frac{\Delta}{\hbar}(\vec{q}-\vec{q'})^2]\fW(\vec{q'},\vec{p'};t).
\eeq
This is known as the Husimi function\cite{Husimi}, 
which is semi-positive definite and is regarded as a quantum distribution function.
Thus we may define the Wehrl entropy in terms of the Husimi function;
\beq
S_{HW}(t)=-\int \frac{d \vec{p} d \vec{q}}{(2 \pi \hbar)^n}\fH(\vec{q},\vec{p};t)\log \fH(\vec{q},\vec{p};t).
\eeq
We call this entropy Husimi-Wehrl(HW) entropy\cite{Wehrl}. 


In the following section, we propose two numerical methods to calculate the time evolution of the HW entropy in the semi-classical approximation.
We first apply these methods to quantum mechanical systems with a few degrees of freedom
and examine their characteristics and usefulness
, which is reported in Ref.\cite{Tsukiji2015}.
Next we apply them to Yang-Mills theory.

\section{The formulation of time evolution and two numerical methods}

{\it Time evolution in semi-classical approximation.}---
We first give the equation of motion for the Wigner function derived from the von Neumann equation;
\beq
\frac{\partial f_W}{\partial t}=\sum^n_i \frac{\partial H}{\partial q_i}\frac{\partial f_W}{\partial p_i}-\sum^n_i \frac{\partial H}{\partial p_i}\frac{\partial f_W}{\partial q_i}+\mathcal{O}(\hbar^2).
\eeq

In the semi-classical approximation, we find that the Wigner function is
constant along the classical trajectory; $\dot{p}_i=\partial H/\partial q_i,\dot{q}_i=-\partial H/\partial p_i$.
Thus we can obtain the time evolution of the Wigner function in the semi-classical approximation by solving the classical equation of motion.
Based on this property,
we propose the following two methods to perform an integration needed for the HW entropy.

{\it Numerical methods.}---
One of the problem to evaluate the HW entropy is that we need integration
over large dimensional space, especially in field theories.
We propose two Monte-Carlo based numerical methods.
The first one is a direct Monte-Carlo integral,
referred to as the two step Monte-Carlo(tsMC) method.
The HW entropy is given in terms of the Wigner function;
\beq
S_{HW}(t)&=&-\int \underline{\frac{d \vec{p} d \vec{q}}{(\pi \hbar)^n}}_{A_I} \underline{\exp[-\frac{1}{\Delta \hbar}\vec{p}^2-\frac{\Delta}{\hbar}\vec{q}^2]}_{A_W} \underline{\int\frac{d \vec{p'}d \vec{q'}}{(2 \pi \hbar)^n}}_{B_I}\underline{\fW(\vec{q'},\vec{p'};t)}_{B_W}\nonumber\\
&\times&\log \underline{\int\frac{d \vec{p''}d \vec{q''}}{(\pi \hbar)^n}}_{C_I}\underline{\exp[-\frac{1}{\Delta \hbar}(\vec{p}+\vec{p'}-\vec{p''})^2-\frac{\Delta}{\hbar}(\vec{q}+\vec{q'}-\vec{q''})^2]}_{C_W}\fW(\vec{q''},\vec{p''};t).~~\label{SHWinW}
\eeq
$A_{I(W)}, B_{I(W)}$ and $C_{I(W)}$ are defined by the underlined integrals(functions) in Eq.~(\ref{SHWinW}). We evaluate the integrals, $A_I, B_I$ and $C_I$, by Monte Carlo method.
For the importance sampling, we generate MC samples according to the weight functions, $A_W, B_W$ and $C_W$.
To prepare the MC samples with the weight of the Wigner function, we generate the samples at $t=0$ according to the initial distribution, and obtain the corresponding phase space sample points at time $t$ by solving the classical equation of motion.
Then the number of the MC samples of integrals, $A_{I(W)}$ and $B_{I(W)}$ ($C_{I(W)})$, out of (in) the logarithmic in Eq.~(\ref{SHWinW}) is described by  ${N}_{\rm in(out)}$.
In this work we assume that the Wigner function is a semi-positive definite function to apply the MC method.

The second one is a test particle(TP) method.
We prepare $N_{\rm TP}$ samples of the (test) particle in phase space 
so that the sample density is proportional to the initial distribution; 
$\fW(\vec{q},\vec{p};t)=\frac{(2 \pi \hbar)^n}{N_{\rm TP}}\sum^{N_{\rm TP}}_i \delta^{(n)}(\vec{p}-\vec{\bar{p}}^i(t))\delta^{(n)}(\vec{q}-\vec{\bar{q}}^i(t))$, where $N_{\rm TP}$ is the total number of the test particles. 
Then the Wigner function is approximately given by the sum of delta functions.
We substitute it into Eq.~(\ref{SHWinW}) and obtain
\beq
S_{HW}(t)&=&-\int \underline{\frac{d \vec{p} d \vec{q}}{(\pi \hbar)^n}}_{D_I}\underline{\exp[-\frac{1}{\Delta \hbar}\vec{p}^2-\frac{\Delta}{\hbar}\vec{q}^2]}_{D_W}\frac{1}{N}\sum^{N_{\rm TP}}_i\nonumber\\
&\times&\log \frac{2^{N_{\rm TP}}}{N}\sum^{N_{\rm TP}}_j\exp[-\frac{1}{\Delta \hbar}(\vec{p}+\vec{\bar{p}}^i(t)-\vec{\bar{p}}^j(t))^2-\frac{\Delta}{\hbar}(\vec{q}+\vec{\bar{q}}^i(t)-\vec{\bar{q}}^j(t))^2].\label{SHWinTP}
\eeq
$D_{I(W)}$ is defined by the underlined integral(function) in Eq.~(\ref{SHWinTP}).We evaluate the integral $D_I$ by the Monte Carlo method with the weight $D_W$.
The number of the MC samples is described by $N_{\rm MC}$.

We have examined these two methods
in some two-dimensional quantum mechanical systems with quartic interaction.
We find that two methods give consistent results
as long as the sampling number is large enough \cite{Tsukiji2015}.

One of the interesting findings is that when the sampling number is not large enough tsMC tends to overestimate the entropy,
while the test particle method tends to underestimate the entropy.
The actual HW entropy should be between the two results,
because, as the number of samples becomes larger, the results of tsMC(TP) method
approach to the converged value from above(below).

\section{Extention to Yang-Mills field theory}

{\it Classical Yang-Mills field theory.}---
Now let us discuss a Yang-Mills field.
We work in the temporal gauge $A_0=0$, then the Hamiltonian in non-compact formalism is given by
\beq
H=\frac{1}{2} \sum_{x,a,i} E^a_i(x)^2 +\frac{1}{4} \sum_{x,a,i,j} F^a_{i j}(x)^2,
\eeq
with $F^a_{i j}=\partial_i A^a_j(x)-\partial_j A^a_i(x) +\sum_{b,c}f^{a b c}A^b_i(x)A^c_j(x)$.

If we assign $(A,E)$ to the phase space coordinate,
it is straightforward to apply the numerical methods developed
in quantum mechanics.

\begin{figure}[b]
\begin{center}
\includegraphics[width=70mm]{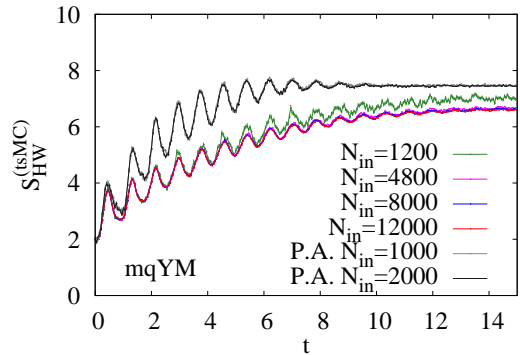}
\caption{The variation in the product ansatz in the two-dimensional quantum model with tsMC.}
\label{2dmqYM-tsMCPA}
\end{center}
\end{figure}

{\it Product ansatz.}---
In the system with a large degrees of freedom, an extremely large number of samples is necessary
for the precision calculation of the HW entropy
even when we adopt the present MC methods.
To avoid the difficulty, we here propose the product ansatz to evaluate the HW entropy
of the Yang-Mills field.
We assume that the Husimi function is written as a product of that for one degree of freedom, $\fH(q,p;t)=\prod^D_i h_i(q_i,p_i;t)$.
Under this ansatz, we can obtain the HW entropy as a sum of the HW entropy for one degree of freedom;
\beq
S_{HW}\simeq -\sum^D_i \int \frac{d q_i d p_i}{2 \pi \hbar}h_i(q_i,p_i;t)\log h_i(q_i,p_i;t).
\eeq

{\it Confirmation in quantum mechanical model.}---
Figure \ref{2dmqYM-tsMCPA} shows that the product ansatz is found to give the HW entropy consistent with the precisely evaluated value
within around 10 \% error in the case of the two-dimensional quantum mechanical problem.


\begin{figure}[t]
\begin{center}
\includegraphics[width=90mm]{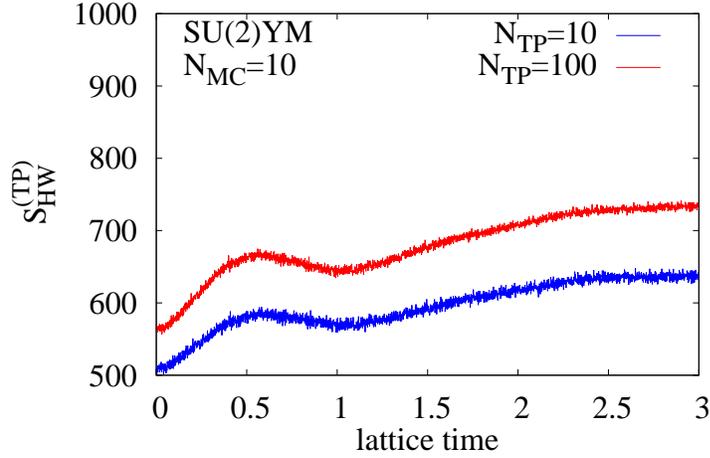}
\caption{Time evolution of Husimi-Wehrl entropy for the SU(2) Yang-Mills field theory on $4^3$ lattice in TP method}
\label{4su2YM-TP}
\end{center}
\end{figure}

\begin{figure}[t]
\begin{center}
\includegraphics[width=90mm]{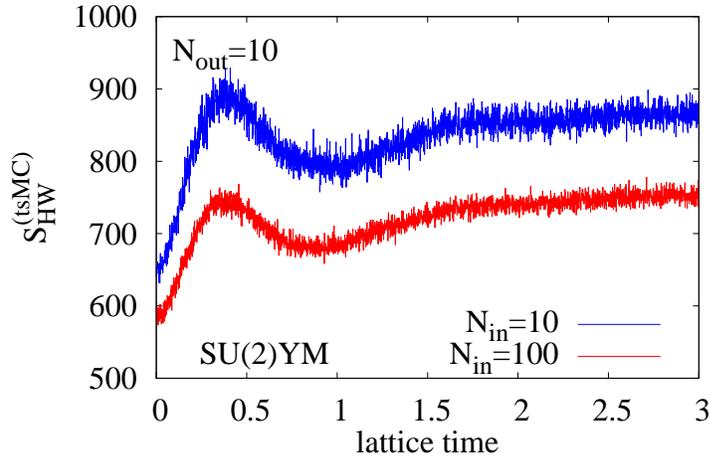}
\caption{Time evolution of Husimi-Wehrl entropy for SU(2) Yang-Mills field theory on $4^3$ lattice in tsMC method}
\label{4su2YM-tsMC}
\end{center}
\end{figure}
 
{\it Results in Yang-Mills field theory.}---
We apply the product ansatz to the color SU(2) Yang-Mills field on $4^3$ and $6^3$ lattices.

Figures~\ref{4su2YM-TP} and \ref{4su2YM-tsMC} show the time dependence of the HW entropy in SU(2) Yang-Mills theory on the $4^3$ lattice with the TP and tsMC methods, respectively.
We find that 
the HW entropy increases as a function of time in both methods,
which implies thermal entropy production. 
As in the quantum mechanics cases,
the HW entropy in the tsMC(TP) method decreases(increases) with increasing Monte-Carlo samples(test particles).
The entropy value at t=3 is almost the same in the two methods with a large number of MC samples(test particles).

\begin{figure}[t]
\begin{center}
\includegraphics[width=90mm]{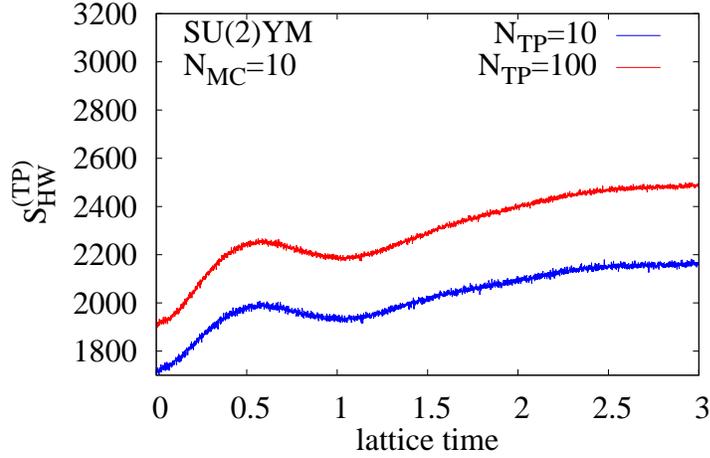}
\caption{Time evolution of Husimi-Wehrl entropy for SU(2) Yang-Mills field theory on $6^3$ lattice in TP method}
\label{6su2YM-TP}
\end{center}
\end{figure}

\begin{figure}[t]
\begin{center}
\includegraphics[width=90mm]{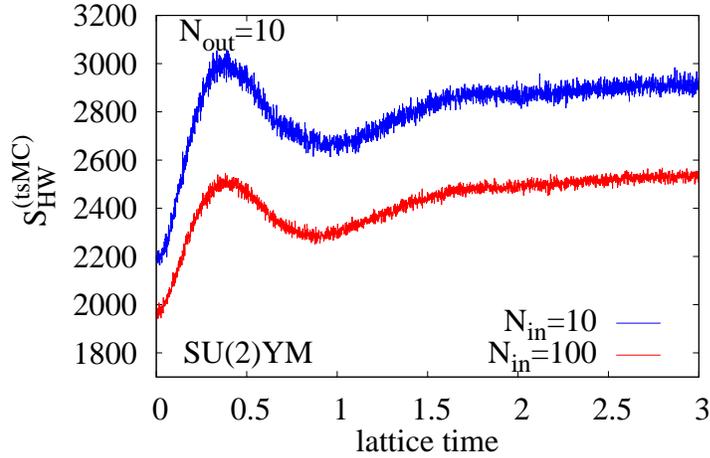}
\caption{Time evolution of Husimi-Wehrl entropy for SU(2) Yang-Mills field theory on $6^3$ lattice in tsMC method}
\label{6su2YM-tsMC}
\end{center}
\end{figure}

Figures~\ref{6su2YM-TP} and \ref{6su2YM-tsMC} are the results in SU(2) Yang-Mills theory on $6^3$ lattice with the TP and tsMC methods, respectively.
The qualitative behavior is the same as that in $4^3$ lattice.
The HW entropy is created also on the larger lattice.

\clearpage
\section{Summary}

%

In order to explore the thermalization process in the early stage of the 
relativistic heavy ion collisions, 
we have proposed to  use a positive semi-definite quantum distribution function called the Husimi function,
 which is given by a Gaussian smearing of the Wigner function within the minimum uncertainty:
 Since the semiclassical approximation is valid in  the system of our interest and the
classical Yang-Mills(YM) dynamics is known to show a chaotic behaior which can be a significant
origin of the entropy production, 
the use of the distribution function is useful because it allows a classical-quantum correspondence.
Then a kind of entropy can be defined in terms of the Husimi function, which is called the Husimi-Wehrl(HW) entropy. 
We have shown that it is feasible to apply the semiclassical approximation to calculate the
HW entropy with the use of the classical YM equation.
We have proposed two numerical methods to calculate the HW entropy on the basis of 
the Monte-Carlo method, which we call two step Monte-Carlo(tsMC) and test particle(TP) methods,
respectively: It is found that the combination of these two methods 
gives reliable results for the entropy production for
 two-dimensional quantum mechanical systems.
We have also introduced a product ansatz for calculating the HW entropy for
a system with large number of degrees of freedom.
We have confirmed that the product ansatz 
roughly reproduce the HW entropy with about 10-20\% overestimate for a few-body quantum system.
We have calculated the HW entropy of (quantum) Yang-Mills field theory with the tsMC and TP methods coupled with product ansatz, and found that
the HW entropy increases in time. This implies that 
thermal entropy can be surely created in the early stage of the relativistic heavy-ion
collisions
, which is triggered
by the initial fluctuations.

\section*{Acknowledgments}
This work is supported in part by KAKENHI
(Nos. 20540265, 
          23340067, 
          24340054, 
          24540271, 
         15H03663, 
         15K05079, 
         24105001, 
and
         24105008), 
and by the Yukawa International Program for Quark-Hadron Sciences.
T.K. is supported by the Core Stage Back Up program in Kyoto University.

\end{document}